\documentclass[showpacs,10pt]{revtex4}
\usepackage{amsmath}
\usepackage{amsfonts}
\usepackage{amssymb}
\usepackage{amsthm}
\usepackage{bm}
\usepackage{mathrsfs}
\usepackage{tensor}
\usepackage{graphicx}

\begin{document}
\title{Generalized Transformation Optics of Linear Materials}
\author{Robert T. Thompson}
\email{robert@cosmos.phy.tufts.edu}
\affiliation{Department of Mathematics and Statistics,
University of Otago, P.O.\ Box 56, Dunedin, 9054,  New Zealand}
\author{Steven A. Cummer}
\affiliation{Department of Electrical and Computer Engineering and Center for Metamaterials
and Integrated Plasmonics, Duke University, Durham, North Carolina 27708, USA}
\author{J\"{o}rg Frauendiener}
\affiliation{Department of Mathematics and Statistics,
University of Otago, P.O.\ Box 56, Dunedin, 9054,  New Zealand}
\affiliation{Centre of Mathematics for Applications, University of Oslo, P.O.\ Box 1053, Blindern, NO-0316 Oslo, Norway}

\begin{abstract}
We continue the development of a manifestly 4-dimensional, completely covariant, approach to transformation optics in linear dielectric materials begun in a previous paper.  
This approach, which generalizes the Plebanski based approach, is systematically applicable for all transformations and all general linear materials.  
Importantly, it enables useful applications such as arbitrary relative motion, transformations from arbitrary non-vacuum initial dielectric media, and arbitrary space-times.  
This approach is demonstrated for a resulting material that moves with uniform linear velocity, and in particular for a moving cloak. 
The inverse problem of this covariant approach is shown to generalize Gordon's ``optical metric''.
\end{abstract}


\maketitle

\section{Introduction}
The emerging field of transformation optics, where useful arrangements of man-made ``metamaterials'' \cite{PhysRevLett.76.2480,Pendry:1999,Smith:2004}  are designed via transformations of electromagnetic fields, has theoretical roots stretching back nearly a century, to the early days of general relativity.  
The idea that the behavior of light in a gravitational field can be replicated by a suitable distribution of refractive media appears to have been first postulated by Eddington \cite{Eddington:1920}. 
Subsequently, Gordon \cite{Gordon:1923} studied the inverse problem, that of representing a refractive medium as a vacuum space-time described by an ``optical metric''.  
Later, Plebanski \cite{Plebanski:1959ff} found effective constitutive relations for electromagnetic waves propagating in vacuum space-times, but, while recognizing the formal equivalence of these equations to those in a macroscopic medium in flat space-time, does not exploit this equivalence to actually describe such a medium.  
This was first done by De Felice \cite{DeFelice:1971}, who used the Plebanski equations to describe the equivalent medium of both a spherically symmetric gravitational system and Friedmann-Robertson-Walker space-time.

More recently, Pendry \cite{Pendry:2006a}, pointed out the specific relationship between spatial transformations and material properties, and demonstrated how it could be used to create novel devices.  
Closely related work by Greenleaf et al. \cite{Greenleaf:2003} developed a similar concept for electric current flow and applied it to impedance tomography.  
The initial approaches to transformation optics relied on purely spatial transformations \cite{Pendry:2006a,Schurig:2006,Milton:2006}. 
Leonhardt and Philbin generalized this to transformations involving both space and time by using the explicit equivalence given by De Felice; thus linking transformation optics to differential geometry \cite{Leonhardt:2006ai}.  Another approach \cite{Tretyakov:2008}, based on field-transforming metamaterials, considers  more general transformations in the Fourier domain, but at the expense of the intuitive and appealing geometric interpretation.
For more recent reviews see Refs.\ \cite{greenleaf:3,Leonhardt:2009}.

In examining the details of the Plebanski-De Felice approach, we find some limitations that are addressed here.  
As pointed out by Plebanski himself \cite{Plebanski:1959ff}, the equation now bearing his name is not strictly covariant, because its derivation requires a matrix inversion that is not a true tensor operation.  
One consequence of this, in the context of transformation optics, is that magneto-electric coupling terms can not always be simply interpreted as a material velocity, as is frequently done.  
This is because the Plebanski equations can only be identified with stationary media or with slowly moving (i.e.\ nonrelativistic), \textit{isotropic} media \cite{Thompson:2011jo}.  
This prompts the question of whether the approach of Ref.\ \cite{Leonhardt:2006ai} may be generalized to clearly distinguish magneto-electric couplings from material velocity, and allow for specially relativistic corrections.
Just such an approach was outlined in Ref.\ \cite{Thompson:2011jo}, and was demonstrated to to recover several results obtained through other means.  

Here we provide a complete derivation of the approach outlined in Ref.\ \cite{Thompson:2011jo}, further generalizing the result obtained there.
A physically realistic scenario for transformation optics designed devices is that the device move with arbitrary velocity. 
We demonstrate that the approach described here may be applied to find the material properties of a transformation when the resultant material is constrained to move with arbitrary uniform velocity with respect to the frame in which the transformations are given.  
This represents a departure from most previous examples in transformation optics, where either the resulting material is stationary or where the velocity is dictated by the transformation itself. 
While a few examples exist of nonrelativistic moving dielectrics in transformation optics for special cases \cite{Cheng:2009,Tretyakov:2008}, we provide a systematic approach that is widely applicable for any velocity. 

Another limitation of the Plebanski-De Felice approach is that the resulting material must reside in vacuum, Minkowski space-time.  
The approach described here relaxes these conditions, allowing for physically realistic scenarios such as transformations in arbitrary non-vacuum initial dielectric media \cite{Thompson:2010pa}, or in arbitrary space-times -- thus providing general relativistic corrections for transformations in arbitrary space-times, such as the weakly curved space-time near Earth \cite{Thompson:2011gr}.  
Lastly, we show that this covariant approach is consistent with, and generalizes, Gordon's optical metric as essentially the inverse problem of transformation optics.

The paper is organized as follows: In Sec.\ \ref{Sec:ClassElec} we review the completely covariant theory of vacuum electrodynamics using modern language; for which the necessary ideas and notation from differential geometry may be found in Appendix \ref{Sec:Geometry}.  
This review is presented in some detail, because in Sec.\ \ref{Sec:LinearMedia} the covariant theory of Sec.\ \ref{Sec:ClassElec} is extended to describe electrodynamics in macroscopic linear dielectric materials.  
This section presents a slight departure from the usual description of electrodynamics in dielectric materials in order to clearly distinguish material effects from space-time effects. 
Section \ref{Sec:TransformationOptics} describes the concept of transformation optics and presents an interpretation consistent with the geometric picture of the preceding sections. 
The main result for applications in transformation optics is Eq.\ (\ref{Eq:MaterialChi}).  
Section \ref{Sec:Examples} examines a particular transformation both when the resulting material is at rest and when it is in motion relative to the frame in which the fields have been measured.  
As expected, it is found that the results for the material in motion smoothly recover, in the limit $\vec{v}\to 0$, the results for the material at rest.  
In Sec.\ \ref{Sec:Gordon} we study the inverse problem of transformation optics, that of finding an equivalent vacuum space-time starting from an initial dielectric, thus generalizing Gordon's optical metric idea.  
We conclude with Sec.\ \ref{Sec:Conclusions}.

\section{Classical Electrodynamics in Vacuum} \label{Sec:ClassElec}
The basic elements of covariant electrodynamics needed for transformation optics were presented in Ref.\ \cite{Thompson:2011jo}, here we present a relatively self-contained and more detailed description of covariant electrodynamics in both vacuum and linear materials.  
This development relies on the geometric language and tools of differential geometry, such as \textit{exterior derivative}, \textit{wedge product}, and the \textit{pullback} of a tensor.  
These aspects of differential geometry are described in a myriad of excellent sources, such as \cite{Tu:2008,Frankel:1997ec,Bleecker,Baez}.  
The most important of these for our purpose, the pullback map, is described in Appendix \ref{Sec:Geometry}. 
The development and notation, in particular the sign convention, follows that of Ref.\ \cite{Misner:1974qy}, while more information, particularly for electrodynamics in materials, can be found in Refs.\ \cite{Post:1962,Hehl}.  
We use the Einstein summation convention, indices are lowered (raised) by the metric tensor $g_{\alpha\beta}$ (its inverse $g^{\alpha\beta}$), and the speed of light and Newton's constant are set to $c=G=1$.
\subsection{Field Strength Tensor}
In free space, classical electrodynamics is modeled as a principal $U(1)$ fiber bundle over a space-time manifold $M$ (which we assume to be equipped with a metric) with connection 1-form $\mathbf{A}=A_{\mu}$ (frequently called a ``gauge field'', $\mathbf{A}$ is the covariant version of the 4-vector potential).  
The \textit{field strength} $\mathbf{F}=F_{\mu\nu}$ is the curvature 2-form of the $U(1)$ fiber bundle, equal to the exterior derivative of $\mathbf{A}$,
\begin{equation}
 \mathbf{F}=\rm{d}\mathbf{A} \Rightarrow F_{\mu\nu} = A_{\nu,\mu}-A_{\mu,\nu}
\end{equation}
where the comma indicates a derivative. 
The components of $\mathbf{F}$ can be represented as a matrix, that in a local orthonormal frame (or Minkowski space-time with Cartesian coordinates) have values
\begin{equation} \label{Eq:FComponents}
 F_{\mu\nu} = 
 \begin{pmatrix}
  0 & -E_x & -E_y & -E_z\\
  E_x & 0 & B_z & -B_y\\
  E_y & -B_z & 0 & B_x\\
  E_z & B_y & -B_x & 0
 \end{pmatrix}.
\end{equation}
With this choice of values for the components of $\mathbf{F}$, the 4-force vector on a particle moving with 4-velocity $u^{\nu}$ and charge $q$ is
\begin{equation}
 f^{\alpha}=q g^{\alpha\mu}F_{\mu\nu}u^{\nu},
\end{equation}
which is just the Lorentz force.  
For example, a particle at rest with respect to this system has 4-velocity $u^{\nu}=(1,0,0,0)$, for which $f^{\mu} = q(0,E_x,E_y,E_z)$, recovering the usual notion that a charged particle at rest feels only the electric part of the field.  
For a particle moving with $u^{\nu}=\gamma(1,v^x,v^y,v^z)$, the spatial components of $f^{\mu}$ are $\vec{f}=q\gamma\left(\vec E + \vec v \times \vec B\right)$, while the time component is the change in energy per unit time, or the power.

The $\vec E$ and $\vec B$ fields are now tightly intertwined, simply representing different components of a single object, $\mathbf{F}$. Because the second exterior derivative of any form vanishes and the fact that $\mathbf{F}=\rm{d}\mathbf{A}$, it immediately follows that
\begin{equation} \label{Eq:HomogeneousMaxwell}
 \rm{d} \mathbf{F} = 0.
\end{equation}
This is nothing more than the covariant form of the homogeneous Maxwell equations, and shows that the homogeneous equations are simply geometric conditions imposed on the fields.

\subsection{Field Strength Dual}
Naturally associated to each point of the space-time manifold are four $m$-dimensional vector spaces, where $m=\mathrm{dim}(M)$.  The metric generates a bijection $g$ between the space of $k$-forms $\wedge^kT_p^*(M)$ and the space of $k$-vectors $\wedge^kT_p(M)$.  The volume form provides a bijection $\omega$ between the space of $k$-forms and the space of $(m-k)$-vectors $\wedge^{(m-k)}T_p(M)$.  The composition of maps, called the Hodge dual, makes the diagram of figure 1 commutative.  The Hodge dual provides a natural two-form dual to the field strength $F_{\mu\nu}$.  In particular, we define the map
\begin{equation}
 \star :\wedge^2T^*_p(M)\to \wedge^2T^*_p(M)
\end{equation}
as the composition $\star = \omega\circ g$ applied to 2-forms (where $\wedge^2T^*_p(M)$ is the space of 2-forms) , or in component form
\begin{equation} \label{Eq:StarF}
 (\star \mathbf{F})_{\mu\nu} = \frac12 \sqrt{|g|}\,\epsilon_{\mu\nu\alpha\beta}g^{\alpha\gamma}g^{\beta\delta}F_{\gamma\delta}.
\end{equation}
The components of $\star \mathbf{F}$ can also be represented as a matrix, that in a local orthonormal frame (or Minkowski space-time with Cartesian coordinates) have values
\begin{equation} \label{Eq:StarFComponents}
 (\star \mathbf{F})_{\mu\nu} = 
 \begin{pmatrix}
  0 & B_x & B_y & B_z\\
  -B_x & 0 & E_z & -E_y\\
  -B_y & -E_z & 0 & E_x\\
  -B_z & E_y & -E_x & 0
 \end{pmatrix}.
\end{equation}
The dual nature is now explicit; where we had the decomposition $\mathbf{F} = E_a \mathrm{d}x^a\wedge \mathrm{d}t + B_{ab}(\mathrm{d}x^a\wedge \mathrm{d}x^b)$ we now have the dual decomposition $\star \mathbf{F} = -B_a \mathrm{d}x^a\wedge \mathrm{d}t + E_{ab}(\mathrm{d}x^a\wedge \mathrm{d}x^b)$.

\begin{figure}[ht]
\includegraphics{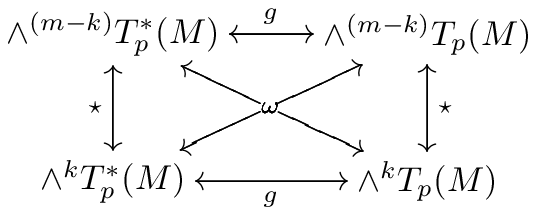}
\caption{The metric generates a bijection $g$ between $\binom{k}{0}$ and $\binom{0}{k}$ alternating tensors, while the volume form $\omega$ provides a bijection between $\binom{k}{0}$ and $\binom{0}{m-k}$ alternating tensors (for any $k< m$).  
The composition $\star = \omega\circ g$ provides a bijection between $\binom{m-k}{0}$ and $\binom{k}{0}$ alternating tensors and between $\binom{0}{m-k}$ and $\binom{0}{k}$ alternating tensors.}  \label{Fig:Diagram}
\end{figure}

\subsection{Vacuum Action}
To describe electrodynamics by means of a variational principle requires an action, and to construct an action $S=\int d^4x \, \sqrt{|g|} \mathcal{L}$ requires a suitable Lagrangian density $\mathcal{L}$.  
Differential geometry tells us that we can only integrate a $k$-form over a $k$-dimensional (sub)manifold.  
There are 3 ways to construct a 4-form from $\mathbf{F}$ and $\mathbf{\star F}$ that can be integrated over the 4-dimensional space-time manifold: $\mathbf{F}\wedge\mathbf{F}$, $\star\mathbf{F}\wedge\star\mathbf{F}$, and $\mathbf{F}\wedge\star\mathbf{F}$ (we neglect a possibility such as $\mathbf{A}\wedge\star\mathbf{A}$, which results in a massive photon described by the Proca equations \cite{Jackson}).  
Consider the first possibility.  Since $\mathbf{F}=\mathrm{d}\mathbf{A}$,
\begin{equation}
 \int_M (\mathbf{F}\wedge\mathbf{F}) = \int_M \mathrm{d}(\mathbf{A}\wedge\mathbf{F}) = \int_{\partial M} (\mathbf{A}\wedge\mathbf{F}).
\end{equation}
Essentially, this term is a total 4-divergence and contributes nothing to the integral.  
Writing out the wedge product shows that $\star\mathbf{F}\wedge\star\mathbf{F}=-\mathbf{F}\wedge\mathbf{F}$, so the second possibility also contributes nothing to the integral.  
The last possibility, however, gives a contribution
\begin{equation} \label{Eq:VacLagrangian}
 \int_M (\mathbf{F}\wedge\star\mathbf{F}) = \int_M d^4x \, \sqrt{|g|}  g(\mathbf{F},\mathbf{F})  = \int_M d^4x \, \sqrt{|g|}  (F^{\mu\nu}F_{\mu\nu}).
\end{equation}
The component form of this expression -- the right hand side of Eq.\ (\ref{Eq:VacLagrangian}) -- is the version most commonly encountered in the literature.  
That this appears in component form as $F^{\mu\nu}F_{\mu\nu}$ should be regarded as a happy coincidence that is an artifact of the classical vacuum; do not loose sight of the fact that the fundamental tensors making up the Lagrangian density are $\mathbf{F}$ and $\star\mathbf{F}$, as this will be generalized shortly.

\subsection{Excitation Tensor}
The field strength tensor $\mathbf{F}$ encodes some information about the fields, namely the electric field strength and the magnetic flux.  
Consider that on the other side of this coin, the magnetic field strength and the electric flux are encoded in another tensor $\mathbf{G}$, called the \textit{excitation tensor}.  
The components of $\mathbf{G}$ can be represented as a matrix, that in a local orthonormal frame (or Minkowski space-time with Cartesian coordinates) have values
\begin{equation} \label{Eq:GComponents}
 G_{\mu\nu} = 
 \begin{pmatrix}
  0 & H_x & H_y & H_z\\
  -H_x & 0 & D_z & -D_y\\
  -H_y & -D_z & 0 & D_x\\
  -H_z & D_y & -D_x & 0
 \end{pmatrix}.
\end{equation}
Then the identification 
\begin{equation}
 \mathbf{G} = \star\mathbf{F}
\end{equation}
is a linear map that takes $\star\mathbf{F}$ to $\mathbf{G}$ and provides a set of constitutive relations for the components of $\mathbf{G}$ in terms of those of $\mathbf{F}$.  
Comparing Eq.\ (\ref{Eq:GComponents}) with Eq.\ (\ref{Eq:FComponents}), it is clear that $\mathbf{G}=\star\mathbf{F}$ reduces to the trivial, vacuum, constitutive relations $H_a=B_a$ and $D_a=E_a$ (where $\varepsilon_0=\mu_0=1$ when $c=1$).  
Thus we find that a trivial linear map recovers the correct constitutive relations in vacuum, but this will be extended to something non-trivial in the next section.

We would like to stress that we take $\mathbf{G}$ to be a tensor, not a tensor density.  
Expressing the excitation as a tensor density is common in the literature, but using $\mathbf{G}=\star\mathbf{F}$ makes use of the metric structure and explicitly shows the space-time contributions encoded by $\star$, which will be very useful when discussing materials and, in particular, transformation optics.

Including an interaction term, the action is generalized to
\begin{equation} \label{Eq:Action}
S = \int \frac12\mathbf{F}\wedge\mathbf{G} + \mathbf{J}\wedge\mathbf{A}.
\end{equation}
In the interaction term, $\mathbf{A}$ is the connection 1-form, and $\mathbf{J} = J_{\alpha\beta\gamma}$ is the charge-current 3-form.  
This action is invariant under a gauge transformation $\mathbf{A}\to \mathbf{A}+\mathrm{d}f$ for some scalar function $f$.  
The 3-form $\mathbf{J}$ is related to the usual 4-vector current $\mathbf{j}=j^{\mu}$ by the volume dual, $\mathbf{J}=\omega(\mathbf{j})$, or in component form
\begin{equation}
 J_{\alpha\beta\gamma} = \sqrt{|g|}\epsilon_{\alpha\beta\gamma\mu}j^{\mu}.
\end{equation}

What advantage is there to thinking in terms of the charge-current 3-form $\mathbf{J}$ rather than the 4-vector $\mathbf{j}$?  
Again, the integral is only defined for forms of the same dimensionality as the manifold on which the integration is performed.  The 3-form $\mathbf{J}$ may be integrated over a three dimensional hypersurface of the space-time manifold.  
Choosing a constant-time hypersurface is equivalent to a spatial 3-volume.  
Integrating $\mathbf{J}$ over this spatial 3-volume gives the charge enclosed.  
Integrating $\mathbf{J}$ over a 1+2 hypersurface corresponding to time and a spatial 2-surface gives the current flowing through the spatial surface.

Equipped with the constraint equation $\rm{d}\mathbf{F} = 0$ (the homogeneous Maxwell equations), and the constitutive Eq.\ (\ref{Eq:Constitutive}), varying the action of Eq.\ (\ref{Eq:Action}) with respect to $\mathbf{A}$ gives
\begin{equation} \label{Eq:InhomogeneousMaxwell}
 \rm{d}\mathbf{G} = \mathbf{J}.
\end{equation}
This comprises the inhomogeneous Maxwell equations, although the reader may be more familiar with the expression obtained by taking the dual of both sides of Eq.\ (\ref{Eq:InhomogeneousMaxwell}).

\section{Electrodynamics in Linear Media} \label{Sec:LinearMedia}
Electrodynamics in materials is somewhat more complicated than electrodynamics in vacuum.  
Here we expand on the brief introduction given in  Ref.\ \cite{Thompson:2011jo}, whose main points are embedded here for a complete and self-contained discussion.  
The microscopic theory of electrodynamics would be a quantum field theory described by some complicated action that includes not only the electromagnetic fields, but also the various matter fields making up the material, along with their interactions and associated gauge fields \cite{Hopfield:1958, Huttner:1992}.  
Considering the vast number of fields contained in a sample of ordinary material, it would be an impossible task to examine the full exact theory.  
Fortunately, in the thermodynamic limit, electrodynamics in media can be described by an effective theory.  
This generally comes in the form of a material dependent set of constitutive relations.

The standard vector relations $\vec{D} = \varepsilon \vec{E}$ and $\vec{H} = \mu^{-1}\vec{B}$ (where $\epsilon$ and $\mu^{-1}$ may be matrix-valued) are frequently combined into an expression such as $ G^{\mu\nu} = \zeta^{\mu\nu\alpha\beta}F_{\alpha\beta}$ \cite{Post:1962,Hehl}. 
Such an expression can be useful, particularly when exploring electrodynamics in the absence of a metric \cite{Obukhov:1999ug,Hehl:2004yk,Hehl:2005hu}.  
However, since we assume the existence of a metric, we instead choose to retain the usual space-time notions of metric and Hodge dual $\star$, and take the minimal approach of extending the trivial constitutive equation $\mathbf{G}=\star\mathbf{F}$ in vacuum to a more general linear constitutive equation \cite{Thompson:2011jo}
\begin{equation} \label{Eq:Constitutive}
 \mathbf{G} = \boldsymbol{\chi}(\star\mathbf{F})
\end{equation}
that in component form reads
\begin{equation} \label{Eq:ConstitutiveIndices}
 G_{\mu\nu} = \chi\indices{_{\mu\nu}^{\alpha\beta}}(\star\mathbf{F})_{\alpha\beta}.
\end{equation}
The tensor $\boldsymbol\chi$ contains information on the dielectric material's properties, and can be thought of as representing an averaging over all the material contributions to an action that describes a more fundamental quantum field theory. 
The motivation for using this constitutive relation is to explicitly separate the space-time effects (i.e.\ the Plebanski relations) from the material effects, which will be useful for transformation optics.

To retain the symmetry properties and usual notions of $\mathbf{G}$ and $\mathbf{F}$, $\boldsymbol{\chi}$ must be independently antisymmetric on its first two and last two indices, and in vacuum $\boldsymbol{\chi}(\star\mathbf{F})=\star\mathbf{F}$.  Thus the classical vacuum is a perfect dielectric for which a trivial $\boldsymbol{\chi}$ describes the electrodynamics, and we extend this idea to a non-trivial $\boldsymbol{\chi}$ describing electrodynamics in arbitrary, linear, dielectric media.  
These conditions reduce the number of free parameters of $\boldsymbol{\chi}$ to 36.  
One may further decompose $\boldsymbol{\chi}$ into principle, skewon, and axion parts \cite{Hehl,Hehl:2007ut}, but we do not consider this here. 
Additional symmetry conditions may be imposed based on thermodynamic or energy conservation arguments, or by the lack of an observed directive effect in naturally occurring stationary materials \cite{Landau:1960,Post:1962}.  
Having recently entered an era of engineered materials which may incorporate active elements \cite{Yuan:2009,Popa:2007}, however, we leave open the discussion of additional symmetries and consider the three conditions above to be the minimal requirements.
The condition $\boldsymbol{\chi}_{vac}(\star\mathbf{F})=\star\mathbf{F}$ is sufficient to uniquely specify all components of $\boldsymbol{\chi}$ for the vacuum, they are \cite{Thompson:2011jo}
\begin{equation} \label{Eq:VacChi}
 \boldsymbol{\chi}_{vac}=(\chi_{vac})_{\gamma\delta}^{\phantom{\gamma\delta}\sigma\rho} = \frac12\left(
\begin{matrix}
 \left(\begin{smallmatrix}
 0 & 0 & 0 & 0\\ 0 & 0 & 0 & 0\\ 0 & 0 & 0 & 0\\ 0 & 0 & 0 & 0\\
 \end{smallmatrix} \right) &
 \left(\begin{smallmatrix}
 0 & 1 & 0 & 0\\ -1 & 0 & 0 & 0\\ 0 & 0 & 0 & 0\\ 0 & 0 & 0 & 0\\
 \end{smallmatrix} \right) &
 \left(\begin{smallmatrix}
 0 & 0 & 1 & 0\\ 0 & 0 & 0 & 0\\ -1 & 0 & 0 & 0\\ 0 & 0 & 0 & 0\\
 \end{smallmatrix} \right) &
 \left(\begin{smallmatrix}
 0 & 0 & 0 & 1\\ 0 & 0 & 0 & 0\\ 0 & 0 & 0 & 0\\ -1 & 0 & 0 & 0\\
 \end{smallmatrix} \right) \\
 \left(\begin{smallmatrix}
 0 & -1 & 0 & 0\\ 1 & 0 & 0 & 0\\ 0 & 0 & 0 & 0\\ 0 & 0 & 0 & 0\\
 \end{smallmatrix} \right) &
 \left(\begin{smallmatrix}
 0 & 0 & 0 & 0\\ 0 & 0 & 0 & 0\\ 0 & 0 & 0 & 0\\ 0 & 0 & 0 & 0\\
 \end{smallmatrix} \right) &
 \left(\begin{smallmatrix}
 0 & 0 & 0 & 0\\ 0 & 0 & 1 & 0\\ 0 & -1 & 0 & 0\\ 0 & 0 & 0 & 0\\
 \end{smallmatrix} \right) &
 \left(\begin{smallmatrix}
 0 & 0 & 0 & 0\\ 0 & 0 & 0 & 1\\ 0 & 0 & 0 & 0\\ 0 & -1 & 0 & 0\\
 \end{smallmatrix} \right) \\
 \left(\begin{smallmatrix}
 0 & 0 & -1 & 0\\ 0 & 0 & 0 & 0\\ 1 & 0 & 0 & 0\\ 0 & 0 & 0 & 0\\
 \end{smallmatrix} \right) &
 \left(\begin{smallmatrix}
 0 & 0 & 0 & 0\\ 0 & 0 & -1 & 0\\ 0 & 1 & 0 & 0\\ 0 & 0 & 0 & 0\\
 \end{smallmatrix} \right) &
 \left(\begin{smallmatrix}
 0 & 0 & 0 & 0\\ 0 & 0 & 0 & 0\\ 0 & 0 & 0 & 0\\ 0 & 0 & 0 & 0\\
 \end{smallmatrix} \right) &
 \left(\begin{smallmatrix}
 0 & 0 & 0 & 0\\ 0 & 0 & 0 & 0\\ 0 & 0 & 0 & 1\\ 0 & 0 & -1 & 0\\
 \end{smallmatrix} \right) \\
 \left(\begin{smallmatrix}
 0 & 0 & 0 & -1\\ 0 & 0 & 0 & 0\\ 0 & 0 & 0 & 0\\ 1 & 0 & 0 & 0\\
 \end{smallmatrix} \right) &
 \left(\begin{smallmatrix}
 0 & 0 & 0 & 0\\ 0 & 0 & 0 & -1\\ 0 & 0 & 0 & 0\\ 0 & 1 & 0 & 0\\
 \end{smallmatrix} \right) &
 \left(\begin{smallmatrix}
 0 & 0 & 0 & 0\\ 0 & 0 & 0 & 0\\ 0 & 0 & 0 & -1\\ 0 & 0 & 1 & 0\\
 \end{smallmatrix} \right) &
 \left(\begin{smallmatrix}
 0 & 0 & 0 & 0\\ 0 & 0 & 0 & 0\\ 0 & 0 & 0 & 0\\ 0 & 0 & 0 & 0\\
 \end{smallmatrix} \right) \\
\end{matrix} \right).
\end{equation}

Equation (\ref{Eq:VacChi}) expresses $\boldsymbol{\chi}$ as a matrix of matrices, the first two indices of $\chi\indices{_{\alpha\beta}^{\mu\nu}}$ give the $\alpha\beta$ component of the large matrix, which is itself a matrix described by the second set of indices.  
The component values of $\boldsymbol{\chi}$ \textit{for the vacuum} are unique and independent of coordinate system. For a more general material, the component values can easily be determined by simply matching the results of the constitutive equation $\mathbf{G}=\boldsymbol{\chi}(\star\mathbf{F})$ with the usual flat-space constitutive relations in a particular coordinate system, as shown in Appendix \ref{Sec:CartesianCoords}.  
The components of the constitutive equation provide a set of six independent equations that can locally be collected in the form
\begin{equation} \label{Eq:ConstitutiveComponents1}
 H_a=(\check{\mu}^{-1})\indices{_a^b}B_b + (\check{\gamma_1}^*)\indices{_a^b}E_b, \quad D_a=(\check{\varepsilon}^*)\indices{_a^b}E_b+ (\check{\gamma_2}^*)\indices{_a^b}B_b,
\end{equation}
where we use the notation $\check{a}$ to denote a $3\times 3$ matrix.  Rearranging these to 
\begin{equation} \label{Eq:ConstitutiveComponents2}
 B_a=(\check{\mu})\indices{_a^b}H_b + (\check{\gamma_1})\indices{_a^b}E_b, \quad D_a=(\check{\varepsilon})\indices{_a^b}E_b+ (\check{\gamma_2})\indices{_a^b}H_b.
\end{equation}
gives the more familiar representation for the constitutive relations.  These three-dimensional representations of the completely covariant Eq.\ (\ref{Eq:Constitutive}) are essentially equivalent, and it is a simple matter to switch between them using
\begin{equation} \label{Eq:ConstitutiveShift}
 \check{\mu}=\left(\check{\mu}^{-1}\right)^{-1}, \quad \check{\varepsilon}=\check{\varepsilon}^*-\check{\gamma_2}^*\check{\mu}\check{\gamma_1}^*, \quad \check{\gamma_1}=-\check{\mu}\check{\gamma_1}^*, \quad \check{\gamma_2} = \check{\gamma_2}^*\check{\mu}.
\end{equation}
One should be aware that these $3\times 3$ matrices are not tensors, but simply components of $\boldsymbol{\chi}$ that have been collected into matrices.  
They could be made into tensors by incorporating the appropriate 3-dimensional Hodge dual of a space-like hypersurface.  
As they stand, Eqs.\ (\ref{Eq:ConstitutiveComponents1}) and (\ref{Eq:ConstitutiveComponents2}) are somewhat misleading, because while $E_a$ and $H_a$ are components of a 1-form, $D_a$ and $B_a$ are really selected components of the 2-forms $D_{ab}dx^a\wedge dx^b$ and $B_{ab}dx^a\wedge dx^b$. 
Thus a more appropriate version of Eqs.\ (\ref{Eq:ConstitutiveComponents2}) would be something like 
\begin{subequations} \label{Eq:ConstitutiveComponents3}
 \begin{equation}
  (\boldsymbol{\star}_{\Sigma}B)_a=(\check{\mu})\indices{_a^b}H_b + (\check{\gamma_1})\indices{_a^b}E_b,
 \end{equation}
 \begin{equation}
  (\boldsymbol{\star}_{\Sigma}D)_a=(\check{\varepsilon})\indices{_a^b}E_b+ (\check{\gamma_2})\indices{_a^b}H_b,
 \end{equation}
\end{subequations}
where $\star_{\Sigma}$ is the Hodge dual on the 3-dimensional space-like hypersurface.  
But this requires that we resolve the space-time into space and time components, selecting an observer to define a direction of time.  
The spatial hypersurface is then orthogonal to the selected direction of time.  
Transforming to the local frame of the selected observer we can make the identifications of Eq.\ (\ref{Eq:CartesianChi}) and then give the constitutive equations a 3-dimensional representation.  
However, this requires greater care, and the inclusion of time transformations is not immediately evident.  
We will therefore continue with a manifestly 4-dimensional approach.

\section{Transformation Optics} \label{Sec:TransformationOptics}
The completely covariant approach to transformation optics was outlined in \cite{Thompson:2011jo}, here we derive this method in greater detail.  
Start with an initial space-time $M$ and field configuration $\left(\mathbf{g},\star,\boldsymbol{\chi},\mathbf{F},\mathbf{G},\mathbf{J}\right)$, where $\mathrm{d}\mathbf{F}=0$, $\mathrm{d}\mathbf{G}=\mathbf{J}$, and $\mathbf{G}=\boldsymbol{\chi}(\star\mathbf{F})$.  
The usual approach to transformation optics begins by imagining a coordinate transformation $T$ that in some way ``deforms'' the manifold.  
For example, in the case of an electromagnetic cloak  \cite{Pendry:2006a,Rahm:200887}, we imagine a coordinate transformation that stretches out a hole in Minkowski space-time (with a point removed).

\begin{figure}[ht]
 \scalebox{0.8}{\includegraphics{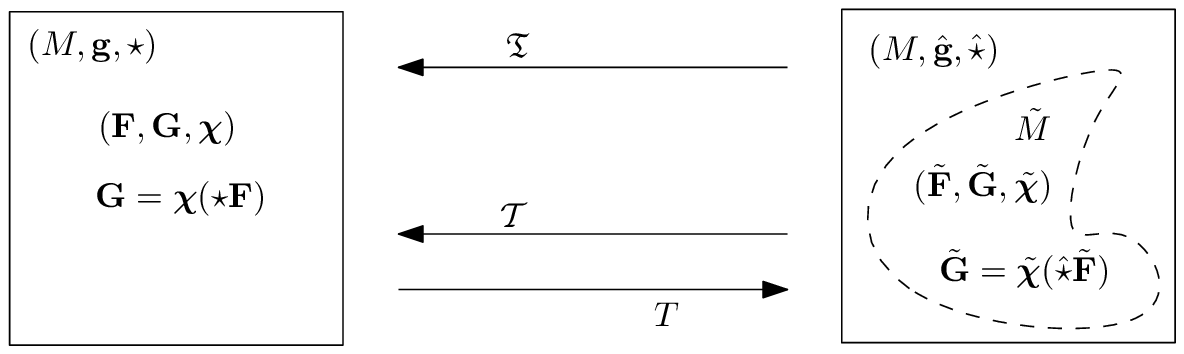}}
 \caption{Under the map $T$ the points of $M$ are mapped to the image $\tilde{M}$.  
 The electromagnetic fields are transformed by the pullback of $\mathcal{T}$, $\mathcal{T}^*$.  
 The metric, on the other hand, is transformed by the pullback of $\mathfrak{T}$, which for present purposes is the identity map.  
 Thus $\tilde{M}\subseteq M$, and the pulled back fields, which exist on $\tilde{M}$, may be excluded from some region of $M$.}
 \label{Fig:Maps}
\end{figure}

However, certain subtleties involved in this picture should be clarified.  
The idea of stretching open a hole or otherwise deforming the space-time is non-physical; therefore it is more appropriate to imagine a map which takes $M$ to an image $\tilde{M}\subseteq M$, as in Fig.\ \ref{Fig:Maps}.  
While it may be intuitive to think of the map $T$ as given, the transformation of 2-forms (representing the fields of interest) requires instead a map $\mathcal{T}$ that describes how the image $\tilde{M}$ is mapped to the original manifold.  
In certain situations it may be possible to set $\mathcal{T}=T^{-1}$, but in general $T^{-1}$ may not even exist.  
Furthermore, while the transformation $\mathcal{T}$ acts to transform the fields, the space-time metric is transformed by a different map $\mathfrak{T}$, which here we take to be the identity.  
We will have more to say about this below.

To make this more concrete, consider a pair of maps $(\mathfrak{T},\mathcal{T})$.  
We demand that $\mathfrak{T}$ acts only on the metric via its pullback
\begin{alignat}{1}
 &\mathfrak{T}:M\to M \\
 &\mathfrak{T}^*(\mathbf{g})=\hat{\mathbf{g}}. \label{Eq:MetricPullback}
\end{alignat}
On the other hand, we demand that the map $\mathcal{T}$ acts on the electromagnetic fields via its pullback
\begin{alignat}{1}
 &\mathcal{T}: \tilde{M}\subseteq M \to M\\
 &\mathcal{T}^*(\mathbf{F}) = \tilde{\mathbf{F}} \label{Eq:FieldPullback}
\end{alignat}
as depicted in Fig.\ \ref{Fig:Maps}.  
Under $(\mathfrak{T},\mathcal{T})$, the metric and fields are transformed from the configuration $\left(\mathbf{g},\star,\boldsymbol{\chi},\mathbf{F},\mathbf{G},\mathbf{J}\right)$, where $\mathrm{d}\mathbf{F}=0$, $\mathrm{d}\mathbf{G}=\mathbf{J}$, and $\mathbf{G}=\boldsymbol{\chi}(\star\mathbf{F})$ to a new configuration $(\hat{\mathbf{g}},\hat{\star},\tilde{\boldsymbol{\chi}},\tilde{\mathbf{F}},\tilde{\mathbf{G}},\tilde{\mathbf{J}})$, where $\mathrm{d}\tilde{\mathbf{F}}=0$, $\mathrm{d}\tilde{\mathbf{G}}=\tilde{\mathbf{J}}$, and $\tilde{\mathbf{G}}= \tilde{\boldsymbol{\chi}}(\hat{\star}\tilde{\mathbf{F}})$.  

The way to physically achieve such a transformation is to introduce some kind of material, just like introducing a dielectric between the plates of a parallel plate capacitor.  
Thus we know that the new configuration of electromagnetic fields must arise from a material distribution $\tilde{\boldsymbol{\chi}}$ and must obey $\tilde{\mathbf{G}}=\tilde{\boldsymbol{\chi}}(\hat{\star}\tilde{\mathbf{F}})$.  
The question of transformation optics is: given a transformation, is it possible to determine the  $\tilde{\boldsymbol{\chi}}$ that will support the transformed fields?

Consider Eqs.\ (\ref{Eq:MetricPullback}) and Eq.\ (\ref{Eq:FieldPullback}) in more detail.  At a point $x\in \tilde{M}$ we have
\begin{equation} \label{Eq:TildeGleft}
 \tilde{\mathbf{G}}_x  = \mathcal{T}^* \left(\mathbf{G}_{\mathcal{T}(x)}\right) 
  =  \mathcal{T}^*\left(\boldsymbol{\chi}_{\mathcal{T}(x)} \circ \star_{\mathcal{T}(x)} \circ \mathbf{F}_{\mathcal{T}(x)}\right),
\end{equation}
where the second line follows from the constitutive relation $\mathbf{G}=\boldsymbol{\chi}(\star\mathbf{F})$.  But from the constitutive relation $\tilde{\mathbf{G}}=\tilde{\boldsymbol{\chi}}(\hat{\star}\tilde{\mathbf{F}})$ we also have
\begin{equation} \label{Eq:TildeGright}
 \tilde{\mathbf{G}}_x=\tilde{\boldsymbol{\chi}}_x\circ \hat{\star}_x \circ \, \mathcal{T}^*\left(\mathbf{F}_{\mathcal{T}(x)}\right).
\end{equation}
The right hand sides of Eqs.\ (\ref{Eq:TildeGleft}) and (\ref{Eq:TildeGright}) must be equal, but to clearly see what is going on we must consider the action of $\tilde{\mathbf{G}}_x$ on a bi-vector $\mathbf{V}_x\in T^2_x(\tilde{M})$, so
\begin{equation} \label{Eq:TransOptics}
 \boldsymbol{\chi}_{\mathcal{T}(x)} \circ \star_{\mathcal{T}(x)} \circ \mathbf{F}_{\mathcal{T}(x)} \circ \mathrm{d}\mathcal{T}(\mathbf{V}_x) = \left[\tilde{\boldsymbol{\chi}}_x\circ \hat{\star}_x \circ \, \mathcal{T}^*\left(\mathbf{F}_{\mathcal{T}(x)}\right)\right]\left(\mathbf{V}_x\right).
\end{equation}
Letting $\Lambda\indices{^{\mu}_{\nu}}$ be the Jacobian matrix of $\mathcal{T}$ ($\Lambda\indices{^{\mu}_{\nu}}$ is the matrix representation of $\mathrm{d}\mathcal{T}$) this may be written in component form as
\begin{equation}
 \chi\indices{_{\alpha\beta}^{\mu\nu}}\Big|_{\mathcal{T}(x)} \star\indices{_{\mu\nu}^{\sigma\rho}}\Big|_{\mathcal{T}(x)} F_{\sigma\rho}\Big|_{\mathcal{T}(x)} \left(\Lambda\indices{^{\alpha}_{\lambda}} \Lambda\indices{^{\beta}_{\kappa}}\right)\Big|_x V^{\lambda\kappa}_x  = \left(\tilde{\chi}\indices{_{\lambda\kappa}^{\xi\zeta}}\, \hat{\star}\indices{_{\xi\zeta}^{\gamma\delta}}\right)\Big|_x F_{\sigma\rho}\Big|_{\mathcal{T}(x)} \left(\Lambda\indices{^{\sigma}_{\gamma}} \Lambda\indices{^{\rho}_{\delta}}\right)\Big|_x V^{\lambda\kappa}_x,
\end{equation}
where we have indicated explicitly where each tensor or Jacobian matrix is evaluated.  
Eliminating $\mathbf{F}$ and $\mathbf{V}_x$ from both sides and solving for $\tilde{\boldsymbol{\chi}}$ gives the final result
\begin{equation} \label{Eq:MaterialChi}
 \tilde{\chi}\indices{_{\lambda\kappa}^{\tau\eta}}(x)=-\Lambda\indices{^{\alpha}_{\lambda}} \Lambda\indices{^{\beta}_{\kappa}} \chi\indices{_{\alpha\beta}^{\mu\nu}}\Big|_{\mathcal{T}(x)} \star\indices{_{\mu\nu}^{\sigma\rho}}\Big|_{\mathcal{T}(x)} (\Lambda^{-1})\indices{^{\pi}_{\sigma}}(\Lambda^{-1})\indices{^{\theta}_{\rho}}\, \hat{\star}\indices{_{\pi\theta}^{\tau\eta}}\Big|_x.
\end{equation}
In Eq.\ (\ref{Eq:MaterialChi}) $\boldsymbol{\Lambda}^{-1}$ is the matrix inverse of $\boldsymbol{\Lambda}$, both $\boldsymbol{\Lambda}$ and $\boldsymbol{\Lambda}^{-1}$ are evaluated at $x$, and in solving for $\tilde{\boldsymbol{\chi}}$ we have made use of the fact that on a 4-dimensional Lorentzian manifold, acting twice with $\star$ returns the negative, $\star\star\mathbf{F}=-\mathbf{F}$.

What is $\hat{\star}$ at the point $x$?  
The Hodge dual $\star$ is not directly transformed by the pullback of $\mathfrak{T}$.  
Rather, the metric is pulled back with $\mathfrak{T}^*$ and then $\hat{\star}$ is computed from the pulled back metric.  
But notice that
\begin{equation}
 \hat{\mathbf{g}}_x = \mathfrak{T}^*(\mathbf{g}_{\mathfrak{T}(x)}) = \mathbf{g}_{\mathfrak{T}(x)}\circ \mathrm{d}\mathfrak{T}
\end{equation}
depends on $\mathbf{g}$ at the point $\mathfrak{T}(x)$, which is not, in general, the same point as $\mathcal{T}(x)$.  
So $\tilde{\boldsymbol{\chi}}(x)$ depends not only on the point $\mathcal{T}(x)$, as explicitly shown in Eq.\ (\ref{Eq:MaterialChi}), but also depends on the point $\mathfrak{T}(x)$ through the evaluation of $\hat{\star}$ at $x$.

Equation (\ref{Eq:MaterialChi}) is the core of transformation optics.  Start with a given space-time with metric $\mathbf{g}$ and associated dual $\star$, and with given dielectric material properties described by the tensor $\boldsymbol{\chi}$.  
The initial space-time may be Minkowski and the initial dielectric may be the vacuum, but this is not necessary.  
We imagine a transformation that changes the fields in some way. We ask what $\tilde{\boldsymbol{\chi}}$ is required to physically achieve such a transformation.  
The answer is given by Eq.\ (\ref{Eq:MaterialChi}).

A few remarks are in order concerning Eq.\ (\ref{Eq:MaterialChi}).  
For applications in transformation optics, $\mathfrak{T}$ is generally taken to be the identity map, meaning that $\mathfrak{T}(x)=x$ and $\hat{\star}=\star$.  
However, this is not strictly necessary, and allowing $\mathfrak{T}$ to be a more general map may be useful in the study of analog space-times \cite{Thompson:2010qh}. 
Furthermore, if in addition to $\mathfrak{T}$ being the identity the initial space-time is vacuum, then since $\boldsymbol{\chi}_{vac}\star = \star$,
\begin{equation} \label{Eq:InitialVacMaterialChi}
 \tilde{\chi}\indices{_{\lambda\kappa}^{\tau\eta}}(x)=-\Lambda\indices{^{\alpha}_{\lambda}} \Lambda\indices{^{\beta}_{\kappa}}  \star\indices{_{\alpha\beta}^{\sigma\rho}}\Big|_{\mathcal{T}(x)} (\Lambda^{-1})\indices{^{\pi}_{\sigma}}(\Lambda^{-1})\indices{^{\theta}_{\rho}}\, \star\indices{_{\pi\theta}^{\tau\eta}},
\end{equation}
where the first $\star$ is evaluated at $\mathcal{T}(x)$ and everything else is evaluated at $x$ \cite{Thompson:2011jo}.  
Notice that the prescriptions of Eqs.\ (\ref{Eq:MaterialChi}) or (\ref{Eq:InitialVacMaterialChi}) are meaningful only for points $x\in\tilde{M}$.  
So for transformations such as the electromagnetic cloak, where there is a hole in $\tilde{M}$, the material parameters inside the hole are unspecified and completely arbitrary.  
In this way, any uncharged material may be hidden inside the cloak without affecting the behavior of the fields outside the cloak.

One might be concerned by the step of assigning two different transformations, one to the metric and one to the fields, and argue that such a step is not allowed because it does not constitute a symmetry of the system.  
However, these are not maps between physically equivalent systems, clearly evidenced by the fact that one is vacuum and the other contains a material.  
Thus there is no symmetry that should be preserved.  
The point of transformation optics is that the transformed fields do not constitute a solution to Maxwell's equations in the original space-time but do constitute a solution to Maxwell's equations in the appropriate material.  
Thus the mappings involved are not mappings between equivalent solutions, but rather a method of generating new solutions from the original (untransformed) solution.

Furthermore, one might be concerned that the pullback of the fields under $\mathcal{T}$ might not be contained in the image of the pullback of the metric under $\mathfrak{T}$.  
Ultimately, this map may only be defined locally. But since the final material parameters must ultimately be measured by a local observer with a local metric, the depiction in Fig.\ \ref{Fig:Maps} can always be defined locally (neglecting irrelevant situations like transformation optics near singularities).

Finally, we note that because of the symmetries of $\mathbf{F}$ and $\mathbf{G}$, which have only 6 independent components each, and of $\boldsymbol{\chi}$ and $\star$, which have at most 36 independent components, we could re-express these results in terms of two 6-dimensional vectors encoding the information of $\mathbf{F}$ and $\mathbf{G}$ and two $6\times 6$ matrices encoding the information of $\boldsymbol{\chi}$ and $\star$.  
This 6-dimensional representation is commonly encountered in the literature, and could be advantageous if performing calculations by hand because of the large number of terms that must be computed in Eq.\ (\ref{Eq:MaterialChi}).  
However, a modern computer algebra package can handle the full calculation with ease, and the manifestly 4-dimensional nature of the expressions is useful when further manipulations are performed.

\subsection{Examples} \label{Sec:Examples}
The completely covariant approach to transformation optics developed above provides a concrete and powerful framework for analyzing any desired configuration of fields and linear dielectric materials in any space-time and with any relative velocities.  
Several previous examples have been given \cite{Thompson:2011jo} which demonstrate that the results obtained with the completely covariant approach agree with results obtained through other means.  
This section presents some additional examples to illustrate the applicability of the completely covariant approach to arbitrary dielectric velocities.

\subsubsection{Mixed Transformation}
Starting from vacuum Minkowski space-time, consider the transformation $T$ defined by
\begin{equation} \label{Eq:MultiCoordTrans}
 T(t', x', y', z')=(t, x, y, z)=(ax't', x', bx'+cy', z')
\end{equation}
The temporal part of this transformation has previously been considered in more detail \cite{Cummer:2011jo}, but the spatial transformation provides some additional complexity.  As discussed above the fields are actually transformed by $\mathcal{T}$ rather than $T$, which we take to be
\begin{equation}
 \mathcal{T}(t,x,y,z) = (t',x',y',z') = \left(\frac{t}{ax},x,\frac{y-bx}{c},z\right), \quad x\neq 0
\end{equation}
and the Jacobian matrix of $\mathcal{T}$ is
\begin{equation}
 \Lambda\indices{^{\mu}_{\nu}} =
 \begin{pmatrix}
  \frac{1}{ax} & -\frac{t}{ax^2} & 0 & 0\\
  0 & 1 & 0 & 0 \\
  0 & -\frac{b}{c} & \frac{1}{c} & 0 \\
  0 & 0 & 0 & 1
 \end{pmatrix}
\end{equation}
Turning the crank on Eq.\ (\ref{Eq:MaterialChi}), and comparing with Eq.\ (\ref{Eq:CartesianChi}), the components of $\boldsymbol{\chi}$ can be extracted and collected as
\begin{equation}
\begin{gathered}
 \check{\varepsilon}^* = \frac{a}{c}x
 \begin{pmatrix}
  1 & b & 0 \\
  b & b^2+c^2-\frac{c^2t^2}{a^2x^4} & 0 \\
  0 & 0 & 1-\frac{t^2}{a^2x^4}
 \end{pmatrix},
\quad
\check{\mu}^{-1} = \frac{1}{acx}
 \begin{pmatrix}
  b^2+c^2 & -b & 0 \\
  -b & 1 & 0 \\
  0 & 0 & c^2
 \end{pmatrix} \\
\check{\gamma_1}^*= -(\check{\gamma_2}^*)^{\mathtt{T}} = \frac{t}{acx^2}
 \begin{pmatrix}
  0 & 0 & b \\
  0 & 0 & -1 \\
  0 & c^2 & 0
 \end{pmatrix}.
\end{gathered}
\end{equation}
Changing to the representation of Eq.\ (\ref{Eq:ConstitutiveComponents2}) by using Eq.\ (\ref{Eq:ConstitutiveShift}) results in
\begin{equation} \label{Eq:IsoSTtrans}
 \check{\varepsilon} = \check{\mu} = \frac{a}{c}x
 \begin{pmatrix}
  1 & b & 0 \\
  b & b^2+c^2 & 0 \\
  0 & 0 & 1
 \end{pmatrix}, \quad
\check{\gamma_1} = \check{\gamma_2}^{\mathtt{T}} = \frac{t}{x}
 \begin{pmatrix}
  0 & 0 & 0 \\
  0 & 0 & 1 \\
  0 & -1 & 0
 \end{pmatrix}.
\end{equation}

The time transformation is responsible for the appearance of the non-zero magneto-electric coupling terms \cite{Thompson:2011jo, Cummer:2011jo}.  
It was shown in Ref.\ \cite{Thompson:2011jo} that if $\check{\varepsilon}=\check{\mu}$ is proportional to the identity matrix, then the magneto-electric coupling terms can be simply identified as a velocity, i.e.\ the magneto-electric coupling may arise solely from a material velocity.  
However, if $\check{\varepsilon}=\check{\mu}$ is not proportional to the identity matrix, as in this example, then the magneto-electric coupling cannot always be interpreted as a simple material velocity.  
Thus in this example we must interpret these results as an anisotropic material with magneto-electric coupling, at rest with respect to the frame in which the fields are given.  
Indeed, it is not possible to boost to a uniformly translating frame in which the magneto-electric coupling vanishes, as we show below.  Note that the time-dependent behavior of these material parameters make a validating calculation quite difficult.  
Such a calculation requires an analysis for both the initial and transformed mediums that is beyond the scope of this paper \cite{Cummer:2011jo}.

\subsubsection{Moving Materials}
In a real world scenario, it is likely that the design parameters require the desired material to move with respect to the frame in which the field transformations are defined.  
In the Plebanski based approach outlined in Ref.\ \cite{Leonhardt:2006ai}, the magneto-electric coupling term is interpreted as a velocity dictated by the transformation, rather than being a free parameter.  
As discussed in Ref.\ \cite{Thompson:2011jo}, a magneto-electric coupling can only be independently identified with a velocity if the material is isotropic.  
Therefore even if the material velocity can be arbitrarily tuned, it is not always possible to tune the velocity such that for a given transformation the material does not require inherent magneto-electric couplings.  
In the totally covariant method presented here, the material velocity is an arbitrary parameter compatible with any value of magneto-electric coupling.

This section presents some examples of a transformation applied to a uniformly moving material in vacuum Minkowski space-time.   
For simplicity, suppose the resulting material moves in the $x$-direction with some speed $\beta$ with respect to the laboratory frame in which the electromagnetic fields are known.  
A prime $v^{\mu'}$ will denote an object in the material frame, while unprimed objects reside in the lab frame.  
A transformation to the material frame from the lab frame is done with a Lorentz boost such that for $v^{\mu'}=L\indices{^{\mu'}_{\nu}}v^{\nu}$,
\begin{equation}
 L\indices{^{\mu'}_{\nu}} =
 \begin{pmatrix}
  \gamma & -\gamma\beta & 0 & 0\\
  -\gamma\beta & \gamma & 0 & 0\\
   0 & 0 & 1 & 0\\
   0 & 0 & 0 & 1
 \end{pmatrix},
\end{equation}
where $\gamma=(1-\beta^2)^{-1/2}$.  The inverse boost, from the material frame to the lab frame, is obtained by setting $\beta\to-\beta$, so that $v^{\nu}=\bar{L}\indices{^{\nu}_{\mu'}}v^{\mu'}$, where
\begin{equation}
 \bar{L}\indices{^{\nu}_{\mu'}} =
 \begin{pmatrix}
  \gamma & \gamma\beta & 0 & 0\\
  \gamma\beta & \gamma & 0 & 0\\
   0 & 0 & 1 & 0\\
   0 & 0 & 0 & 1
 \end{pmatrix}.
\end{equation}
An object with lowered indices boosts as $\omega_{\mu'} = \omega_{\nu}\bar{L}\indices{^{\nu}_{\mu'}}$ or $\omega_{\nu} = \omega_{\mu'}L\indices{^{\mu'}_{\nu}}$.

The calculation now follows more or less straightforwardly from Eq.\ (\ref{Eq:InitialVacMaterialChi}), but includes a boost.  
Let $\boldsymbol{\chi}'$ be the material parameters in the rest frame of the material, the material parameters described in the lab frame are, schematically
\begin{equation}
 \boldsymbol{\chi}_{\mathrm{lab}}=\mathbf{L}^{-1}\left(\boldsymbol{\chi}'\right).
\end{equation}
But the material parameters in the lab frame are precisely those obtained from Eq.\ (\ref{Eq:InitialVacMaterialChi}), $\boldsymbol{\chi}_{\mathrm{lab}}=\tilde{\boldsymbol{\chi}}$.  
It follows that in the frame of the material, where the material parameters are to be measured
\begin{equation} \label{Eq:BoostedMaterialChi}
 \chi\indices{_{\varphi'\xi'}^{\psi'\zeta'}} = - \bar{L}\indices{^{\lambda}_{\varphi'}}\bar{L}\indices{^{\kappa}_{\xi'}}L\indices{^{\psi'}_{\tau}}L\indices{^{\zeta'}_{\eta}}\left(\Lambda\indices{^{\alpha}_{\lambda}} \Lambda\indices{^{\beta}_{\kappa}}  \star\indices{_{\alpha\beta}^{\sigma\rho}} (\Lambda^{-1})\indices{^{\pi}_{\sigma}}(\Lambda^{-1})\indices{^{\theta}_{\rho}}\, \star\indices{_{\pi\theta}^{\tau\eta}},\right).
\end{equation}
Let the transformation again be that of Eq.\ (\ref{Eq:MultiCoordTrans}).  Performing the calculation, extracting the components of $\check{\varepsilon}^*$, $\check{\mu}^{-1}$, $\check{\gamma_1}^*$ and $\check{\gamma_2}^*$, and converting to the representation of Eq.\ (\ref{Eq:ConstitutiveComponents2}), results in
\begin{equation} \label{Eq:BoostedResults1}
 \begin{gathered}
   \check{\varepsilon} = \check{\mu} = \frac{acx^3}{a^2x^4(b^2+c^2)\beta^2-c^2\left(x+t\beta\right)^2}
 \begin{pmatrix}
  a^2x^2\beta^2-\left(1+\frac{t}{x}\beta\right)^2 & -\frac{b}{\gamma}\left(1+\frac{t}{x}\beta\right) & 0 \\
  -\frac{b}{\gamma}\left(1+\frac{t}{x}\beta\right) & -\frac{b^2+c^2}{\gamma^2} & 0 \\
  0 & 0 & -\frac{1}{\gamma^2} 
 \end{pmatrix},
\\
\check{\gamma_1}=\check{\gamma_2}^{\mathtt{T}}= \frac{1}{a^2x^4(b^2+c^2)\beta^2-c^2\left(x+t\beta\right)^2}
 \begin{pmatrix}
  0 & 0 & * \\
  0 & 0 & * \\
  -\frac{ba^2x^4}{\gamma}\beta & c^2(x+t\beta)(t+x\beta)-a^2x^4\beta(b^2+c^2) & 0
 \end{pmatrix},
 \end{gathered}
\end{equation}
where $*$ indicates components that are obtained by the antisymmetry of the matrix.

While these results are fairly complicated and perhaps do not offer much intuition regarding moving dielectrics, this example demonstrates the method by which material velocities may be easily incorporated into the theory, even for non-trivial transformations.  
Essentially, the calculation consists of calculating $\tilde{\boldsymbol{\chi}}$ in the frame in which the desired transformation is specified, and then transforming that result to the rest frame of the corresponding moving material. 
It can be readily seen that the limit $\beta\to 0$ recovers the results of Eq.\ (\ref{Eq:IsoSTtrans}).

As a somewhat simpler example, consider the transformation
\begin{equation}
 \mathcal{T}(t,x,y,z)=(t',x',y',z')=\left(at+bx,x,\frac{y-cx}{d},z\right),
\end{equation}
which is similar to that of Eq.\ (\ref{Eq:MultiCoordTrans}), but avoids the complication of time dependent parameters found in Eqs.\ (\ref{Eq:IsoSTtrans}) and (\ref{Eq:BoostedResults1}).  For a stationary material, the equivalent material parameters are
\begin{subequations} \label{Eq:Multi2Results}
\begin{equation} \label{Eq:Multi2Results1}
 \check{\varepsilon} = \check{\mu} = \frac{1}{ad}
   \begin{pmatrix}
    1 & c & 0\\
    c & c^2+d^2 & 0\\
    0 & 0 & 1
   \end{pmatrix},
\end{equation}
\begin{equation} \label{Eq:Multi2Results2}
 \check{\gamma_1}=\check{\gamma_2}^{\mathtt{T}}= \frac{b}{a}
   \begin{pmatrix}
    0 & 0 & 0\\
    0 & 0 & 1\\
    0 & -1 & 0
   \end{pmatrix}.
\end{equation}
\end{subequations}
It is tempting to associate the magneto-electric coupling with a simple material velocity in the $x$-direction, proportional to $b/a$.
To check whether this is the case, suppose we take a material described by Eq.\ (\ref{Eq:Multi2Results1}) for $\check{\varepsilon}=\check{\mu}$, but with $\gamma_1=\gamma_2=0$, and  moving in the $x$-direction with speed $\beta$.  
In general, $\check{\varepsilon}=\check{\mu}$, and $\check{\gamma_1}=\check{\gamma_2}$  will all now depend on $\beta$.  For low speeds, expanding to leading order in $\beta$ finds $\check{\varepsilon}=\check{\mu}$ unchanged to first order, but with
\begin{equation}
 \check{\gamma_1}=\check{\gamma_2}^{\mathtt{T}}= \frac{\beta}{a^2d^2}
   \begin{pmatrix}
    0 & 0 & -c\\
    0 & 0 & (a^2d^2-c^2-d^2)\\
    c & -(a^2d^2-c^2-d^2) & 0
   \end{pmatrix}.
\end{equation}

Thus interpreting the magneto-electric coupling of Eq.\ (\ref{Eq:Multi2Results2}) as a simple low-speed motion in the $x$-direction is incorrect, as this generates additional magneto-electric coupling components of the same order.  
However, $\check{\gamma_1}_{xz}=c\beta/a^2d^2$ indicates that including a $y$-component for the velocity might allow us to recover the desired results.  
Indeed this is the case, and it can be shown that a material with $\check{\varepsilon}=\check{\mu}$ described by Eq.\ (\ref{Eq:Multi2Results1}) and $\gamma_1=\gamma_2=0$, moving with velocity components
\begin{equation}
 \begin{gathered}
  \beta_x=\frac{ab(a^2f^2-1)}{1-a^2(1-c^2)-f^2(1-a^2)},\\
  \beta_y=\frac{abc}{1-a^2(1-c^2)-f^2(1-a^2)},
 \end{gathered}
\end{equation}
generates the results of Eqs.\ (\ref{Eq:Multi2Results}) to leading order in $\beta_x$ and $\beta_y$.  Velocity induced corrections to $\check{\varepsilon}=\check{\mu}$ are of second order, i.e.\ proportional to $\beta_x\beta_y$, and of course we are restricted to $(\beta_x^2+\beta_y^2)^{1/2} = |\beta|<< 1$.

\subsubsection{Moving cloak}
Finally, consider a cloak moving in the $x$-direction with constant speed $\beta$.  What velocity induced corrections are required to achieve cloaking with respect to fields described in the laboratory frame?  A square cloak is obtained from the transformation \cite{Rahm:200887}
\begin{equation}
  \mathcal{T}(t,x,y,z)=(t',x',y',z')=\left(t,\frac{s_2(x-s_1)}{s_2-s_1},\frac{s_2y(x-s_1)}{x(s_2-s_1)},z\right),
\end{equation}
for $(s_1\leq x\leq s_2)$, $(-s_2< y \leq s_2)$, and $|y|<|x|$.  From Eq.\ (\ref{Eq:InitialVacMaterialChi}), the well-known results for the equivalent material parameters of a stationary cloak are
\begin{equation}
 \check{\varepsilon}=\check{\mu} = 
  \begin{pmatrix}
   1-\frac{s_1}{x} & -s_1\frac{y}{x^2} & 0\\
   -s_1\frac{y}{x^2} & \frac{x^4+s_1^2y^2}{x^3(x-s_1)} & 0\\
   0 &  0 & \frac{s_2^2(x-s_1)}{x(s_1-s_2)^2},
  \end{pmatrix}
\end{equation}
with $\gamma_1=\gamma_2=0$.  Now assume that the cloak is moving.  From Eq.\ (\ref{Eq:BoostedMaterialChi}), it follows that in the frame of the cloak and to first order in the speed $\beta$, the material parameters now have the additional magneto-electric couplings
\begin{equation}
 \check{\gamma_1}=\check{\gamma_2}^{\mathtt{T}}= \frac{\beta s_1}{x^3(s_1-s_2)^2}
   \begin{pmatrix}
    0 & 0 & y(x-s_1)s_2^2\\
    0 & 0 & -x^3(2s_2-s_1)-\frac{y^2}{x}s_1s_2^2\\
   - y(x-s_1)s_2^2 & x^3(2s_2-s_1)+\frac{y^2}{x}s_1s_2^2 & 0
   \end{pmatrix}.
\end{equation}
The moving cloak therefore requires inherent magneto-electric couplings designed to compensate for these velocity induced magneto-electric couplings, which are more complicated than might be expected from previous experience with isotropic materials.  These corrections are likely to be small for ordinary non-relativistic velocities, but depend specifically on the other cloak parameters.  As before, the corrections to $\check{\varepsilon}=\check{\mu}$ are second order in $\beta$.

\section{The Gordon Formalism} \label{Sec:Gordon}
In a problem closely related to the methods of transformation optics, Gordon, in 1923, studied the possibility of representing electrodynamics in a material by an equivalent vacuum space-time \cite{Gordon:1923}.  
One motivation for such a representation is that it allows the material to be described by the action of Eq.\ (\ref{Eq:Action}), with the constitutive relation $\mathbf{G}=\star\mathbf{F}$, where $\star$ is the Hodge dual for a vacuum space-time described by an effective, or ``optical'' metric.  
This is, in some sense, the ``inverse'' problem of transformation optics.  
In particular, suppose we are given a dielectric material described by $\boldsymbol{\chi}$ residing in a space-time with metric $\mathbf{g}$ and corresponding dual $\star$, then from Eq.\ (\ref{Eq:TransOptics}) it may be possible to find $\hat{\star}$ corresponding to an effective vacuum space-time,
\begin{equation} \label{Eq:Gordon}
 \boldsymbol{\chi}\star = \boldsymbol{\chi}_{\mathrm{vac}}\hat{\star} = \hat{\star}.
\end{equation}
To be more explicit, the initial setup consists of a material residing in a space-time, for which the relevant part of the action is
\begin{equation}
 S = \int_M \mathbf{F}\wedge \mathbf{G} = \int_M \mathbf{F}\wedge \boldsymbol{\chi}(\star\mathbf{F}).
\end{equation}
Similarly to Eq.\ (\ref{Eq:VacLagrangian}), this is
\begin{equation}
 \int_M \mathbf{F}\wedge \boldsymbol{\chi}(\star\mathbf{F}) = \int_M  d^4x \sqrt{|g|} \, g(\mathbf{F},\text{-}(\star\,\boldsymbol{\chi}\star\mathbf{F})).
\end{equation}
If this is to be regarded as a vacuum space-time with metric $\boldsymbol{\gamma}$, then
\begin{equation}
  \int_M \mathbf{F}\wedge \boldsymbol{\chi}(\star\mathbf{F}) = \int_M \mathbf{F}\wedge \hat{\star}\mathbf{F} = \int_M  d^4x \sqrt{|\gamma|} \, \gamma(\mathbf{F}, \mathbf{F}).
\end{equation}

The most general $\boldsymbol{\chi}$ has 36 independent components, but the metric only has 10 components, meaning that not every dielectric material can be represented by an equivalent space-time.
To find the conditions that a material must satisfy to be representable as a vacuum space-time, return first to Eq.\ (\ref{Eq:Gordon}).  
Calculating $\hat{\star}$ for an arbitrary metric $\boldsymbol{\gamma}$, and comparing with $\boldsymbol{\chi}\star$ for arbitrary $\boldsymbol{\chi}$, it is readily observed that both $\check{\varepsilon}^*$ and $\check{\mu}^{-1}$ must be symmetric, while $\check{\gamma_2}^*=-(\check{\gamma_1}^*)^{\mathtt{T}}$ is traceless.  
This reduces the number of apparently independent equations to 21, plus one constraint.  
However, the equations are not linear and are given in a particular representation.  Converting to the representation of Eq.\ (\ref{Eq:ConstitutiveComponents2}) one finds that  $\check{\varepsilon}=\check{\mu}$, and that $\check{\gamma_2}=\check{\gamma_1}^{\mathtt{T}}$ is not just traceless but antisymmetric.  
These, of course, are precisely the constraints expected from the Plebanski-De Felice equations.

For a dielectric material to be representable by a curved vacuum space-time it must have $\check{\varepsilon}=\check{\mu}$, and $\check{\gamma_2}=\check{\gamma_1}^{\mathtt{T}}$, which reduces the number of free parameters of $\boldsymbol{\chi}$ to 9.  
Unfortunately, this is insufficient to uniquely determine the 10 free parameters of the metric.  
On a four dimensional Lorentzian manifold, every metric $\mathbf{g}$ is uniquely associated with a map $\star$ from 2-forms to 2-forms by
\begin{equation}
 \star\indices{_{\alpha\beta}^{\mu\nu}} = \frac12 \sqrt{|g|}\epsilon_{\alpha\beta\sigma\rho}g^{\sigma\mu}g^{\rho\nu}.
\end{equation}
Notice that $\star$ is invariant to a scaling of the metric $\mathbf{g} \to a\mathbf{g}$, so given $\star$ it is only possible to determine the metric up to an overall scale factor.  
On the other hand this means that when trying to determine a particular $\boldsymbol{\gamma}$ from a given $\hat{\star}$, one is free to choose a convenient representative from the equivalence class of conformally related space-times, such as the representative with $\gamma_{00}=-1$.  
The situation may improve somewhat if there are sources present, $\mathbf{J}\neq 0$, since the Hodge dual of 1-forms and 3-forms is not invariant under conformal transformations of the metric.  
This may impose some additional constrains on the metric, but remains to be fully investigated.

There exists another condition from which to obtain constraints on the material parameters.  
Recall that one of the defining characteristics of the Hodge dual on a 4-dimensional Lorentzian manifold is that $\star\star\mathbf{F} =-\mathbf{F}$. 
It follows that
\begin{equation}
 \left(\boldsymbol{\chi}\star\right)\left(\boldsymbol{\chi}\star\right) = -id_2
\end{equation}
where we write $id_2$ as the identity of the map from 2-forms to 2-forms.  
This does not provide any independent constraints on the material, but may serve as a useful check when performing calculations.

It is not our purpose to find solutions for a particular distribution of materials, and it is sufficient to remark that, for the case of an isotropic material with vanishing magneto-electric coupling, it is relatively easy to check that Gordon's optical metric
 \begin{equation}
 \gamma_{\mu\nu} = g_{\mu\nu} + \left(1-\varepsilon\mu\right)u_{\mu}u_{\nu}, \quad
  \gamma^{\mu\nu} = g^{\mu\nu} - \left(\varepsilon\mu-1\right)u^{\mu}u^{\nu},
 \end{equation}
where $u^{\mu}$ is the 4-velocity of the material, agrees with our method.  
For an observer at rest with the material, $u^{\mu}=(1,0,0,0)$, while for a material in relative motion to the observer, the $\boldsymbol{\chi}$ in Eq.\ (\ref{Eq:Gordon}) is first transformed with the appropriate boost.

\section{Conclusions} \label{Sec:Conclusions}
We have continued the development of a fully covariant and manifestly 4-dimensional formalism for transformation optics in general linear materials that began in Ref.\ \cite{Thompson:2011jo}.  
The main result of this construction, Eq.\ (\ref{Eq:MaterialChi}), represents a sort of master equation for transformation optics of general, linear, nondispersive media.  
The benefit of this approach is that it is valid for arbitrary background space-times \cite{Thompson:2011gr}, arbitrary initial material \cite{Thompson:2010pa}, and for arbitrary transformations.  
The motivation for this construction was the observation that the Plebanski-De Felice equations that are frequently used in transformation optics and identified with material velocities are only valid in the low velocity limit for isotropic materials.

To arrive at the completely covariant method it was first necessary to find a modest generalization of electrodynamics in linear dielectric media.  
This came by way of a modification to the constitutive relations to the form $\mathbf{G}=\boldsymbol{\chi}(\star\mathbf{F})$.  
Within these relations the vacuum is identified as being a perfect linear dielectric with relative permittivity and permeability equal to 1 -- corresponding to a trivial $\boldsymbol{\chi}$ -- and is therefore treated just like any other dielectric media.

Next, the interpretation of transformation optics was clarified by realizing that the system is being transformed by two distinct maps: one applied to the metric and a different one applied to the electromagnetic fields; because a transformation of the fields is physically realized by changing the dielectric material properties (e.g.\ by introducing a non-vacuum dielectric material to a vacuum region) but generally not the metric.

As an illustration of the completely covariant approach we have examined the results of a transformation when the resulting material is constrained to move with arbitrary uniform velocity -- a physically realistic scenario that has not, as far as we know, been previously considered.   
This is a departure from previous results, where the material velocity has been dictated by the transformation or has been constrained to the nonrelativistic limit. 
It is furthermore shown that the magneto-electric coupling does not arise as straightforwardly from a material velocity as expected from previous interpretations of such couplings.  
For the particular transformation considered it was found that it is in fact possible to approximately get magneto-electric couplings from a material velocity, but only in the low velocity limit where corrections to the permeability and permittivity are neglected, and even then the velocity was in an unexpected direction.
It was shown that for a moving cloak, magneto-electric couplings on the order of the velocity have to be designed into the cloak to compensate for the velocity-induced couplings.

It was shown in Sec.\ \ref{Sec:Gordon} that our formalism generalizes, in a covariant way, Gordon's optical metric \cite{Gordon:1923}, which turns out to be essentially the inverse problem of transformation optics.  
Namely, given a dielectric, can we find an equivalent vacuum space-time?  
The answer appears to be yes, but only if the dielectric satisfies certain properties, and even then only up to a conformal factor of the metric.

It is worth reiterating that the component form of the constitutive relations in the covariant formalism are of the form of Eq.\ (\ref{Eq:ConstitutiveComponents1}) rather than Eq.\ (\ref{Eq:ConstitutiveComponents2}).  
This leads to a slightly different meaning for the $3\times 3$ matrices $\check{\varepsilon}^*$, $\check{\mu}^{-1}$, $\check{\gamma_1}^*$, and $\check{\gamma_2}^*$ than what would be expected from the relations Eq.\ (\ref{Eq:ConstitutiveComponents2}). 
One may readily switch back and forth between the two representations for the constitutive relations via Eq.\ (\ref{Eq:ConstitutiveShift}).  
Keep in mind, however, that the $3\times 3$ matrices of Eq.\ (\ref{Eq:ConstitutiveComponents1}) are merely selected components of the main quantity of interest, the covariant $\binom{2}{2}$-tensor $\boldsymbol{\chi}$. 

\begin{acknowledgments}
The Duke University component of this work was partly supported by a Lockheed Martin University Research Initiative award and by a Multiple University Research Initiative from the Army Research Office (Contract No. W911NF-09-1-0539).
\end{acknowledgments}

\appendix

\section{Cartesian Coordinates} \label{Sec:CartesianCoords}
In Cartesian coordinates, the field strength tensor is defined as
\begin{equation}
 \mathbf{F} = E_x \mathrm{d}x\wedge\mathrm{d}t + E_y\mathrm{d}y\wedge\mathrm{d}t +E_z\mathrm{d}z\wedge\mathrm{d}t + B_x\mathrm{d}y\wedge\mathrm{d}z - B_y\mathrm{d}x\wedge\mathrm{d}z +  B_z\mathrm{d}x\wedge\mathrm{d}y.
\end{equation}
This may be represented in matrix form by Eq.\ (\ref{Eq:FComponents}).
To recover the usual component relations of Eq.\ (\ref{Eq:ConstitutiveComponents1}) the constitutive equation $\mathbf{G}=\boldsymbol{\chi}(\star\mathbf{F})$ allows us to make the identification \cite{Thompson:2011jo}
\begin{equation} \label{Eq:CartesianChi}
 \chi\indices{_{\gamma\delta}^{\sigma\rho}} = \frac12\left(
\begin{matrix}
 \left(\begin{smallmatrix}
 0 & 0 & 0 & 0\\ 0 & 0 & 0 & 0\\ 0 & 0 & 0 & 0\\ 0 & 0 & 0 & 0\\
 \end{smallmatrix} \right) &

* &

* &

* \\[15pt]

 \left(\begin{smallmatrix}
 0 & -\mu^{-1}_{xx} & -\mu^{-1}_{xy} & -\mu^{-1}_{xz}\\ 
 \mu^{-1}_{xx} & 0 & -\gamma_{1xz} & \gamma_{1xy}\\ 
 \mu^{-1}_{xy} & \gamma_{1xz} & 0 & -\gamma_{1xx}\\ 
 \mu^{-1}_{xz} & -\gamma_{1xy} & \gamma_{1xx} & 0
 \end{smallmatrix} \right) &

 \left(\begin{smallmatrix}
 0 & 0 & 0 & 0\\ 0 & 0 & 0 & 0\\ 0 & 0 & 0 & 0\\ 0 & 0 & 0 & 0
 \end{smallmatrix} \right) &

* &

* \\[20pt]

 \left(\begin{smallmatrix}
 0 & -\mu^{-1}_{yx} & -\mu^{-1}_{yy} & -\mu^{-1}_{yz}\\ 
 \mu^{-1}_{yx} & 0 & -\gamma_{1yz} & \gamma_{1yy}\\ 
 \mu^{-1}_{yy} & \gamma_{1yz} & 0 & -\gamma_{1yx}\\ 
 \mu^{-1}_{yz} & -\gamma_{1yy} & \gamma_{1yx} & 0
 \end{smallmatrix} \right) &

 \left(\begin{smallmatrix}
 0 & -\gamma_{2zx} & -\gamma_{2zy} & -\gamma_{2zz}\\
 \gamma_{2zx} & 0 & -\epsilon_{zz} & \epsilon_{zy}\\
 \gamma_{2zy} & \epsilon_{zz} & 0 & -\epsilon_{zx}\\
 \gamma_{2zz} & -\epsilon_{zy} & \epsilon_{zx} & 0
 \end{smallmatrix} \right) &

 \left(\begin{smallmatrix}
 0 & 0 & 0 & 0\\ 0 & 0 & 0 & 0\\ 0 & 0 & 0 & 0\\ 0 & 0 & 0 & 0\\
 \end{smallmatrix} \right) &

* \\[20pt]

 \left(\begin{smallmatrix}
 0 & -\mu^{-1}_{zx} & -\mu^{-1}_{zy} & -\mu^{-1}_{zz}\\ 
 \mu^{-1}_{zx} & 0 & -\gamma_{1zz} & \gamma_{1zy}\\ 
 \mu^{-1}_{zy} & \gamma_{1zz} & 0 & -\gamma_{1zx}\\ 
 \mu^{-1}_{zz} & -\gamma_{1zy} & \gamma_{1zx} & 0
 \end{smallmatrix} \right) &

 \left(\begin{smallmatrix}
 0 & \gamma_{2yx} & \gamma_{2yy} & \gamma_{2yz}\\
 -\gamma_{2yx} & 0 & \epsilon_{yz} & -\epsilon_{yy}\\
 -\gamma_{2yy} & -\epsilon_{yz} & 0 & \epsilon_{yx}\\
 -\gamma_{2yz} & \epsilon_{yy} & -\epsilon_{yx} & 0
 \end{smallmatrix} \right) &

 \left(\begin{smallmatrix}
 0 & -\gamma_{2xx} & -\gamma_{2xy} & -\gamma_{2xz}\\
 \gamma_{2xx} & 0 & -\epsilon_{xz} & \epsilon_{xy}\\
 \gamma_{2xy} & \epsilon_{xz} & 0 & -\epsilon_{xx}\\
 \gamma_{2xz} & -\epsilon_{xy} & \epsilon_{xx} & 0
 \end{smallmatrix} \right) &

 \left(\begin{smallmatrix}
 0 & 0 & 0 & 0\\ 0 & 0 & 0 & 0\\ 0 & 0 & 0 & 0\\ 0 & 0 & 0 & 0
 \end{smallmatrix} \right) \\
\end{matrix} \right).
\end{equation}
where the $*$ indicates entries that are antisymmetric on either the first or second set of indices on $\chi\indices{_{\gamma\delta}^{\sigma\rho}}$.

In cylindrical coordinates $(t,r,\theta,z)$, the Minkowski metric tensor is $g_{\mu\nu} = diag(-1,1,r^2,1)$. 
Because we have $c=1$, time and space are measured in the same units; meaning that $E^{\alpha}$ and $B^{\alpha}$ are both measured in units of $V/m$ and the units for $\mathbf{F}$ are $V\cdot m$.  
Changing to cylindrical coordinates does not change the units of $\mathbf{F}$, so
\begin{equation}
 \mathbf{F} = E_r \mathrm{d}r\wedge\mathrm{d}t + r E_{\theta}\mathrm{d}\theta\wedge\mathrm{d}t +E_z\mathrm{d}z\wedge\mathrm{d}t + r B_r\mathrm{d}\theta\wedge\mathrm{d}z - B_{\theta}\mathrm{d}r\wedge\mathrm{d}z +  r B_z\mathrm{d}r\wedge\mathrm{d}\theta.
\end{equation}
This definition for the components of $\mathbf{F}$ in cylindrical coordinates assumes that all $E^{\alpha}$ and $B^{\alpha}$ are still measured in units of $V/m$, despite the dimensionless coordinate $\theta$.  
Similar arguments hold for $\mathbf{G}$.  These are written in matrix form as
\begin{equation} \label{Eq:CylindricalF}
 F_{\mu\nu} = \left(
 \begin{matrix}
  0 & -E_r & -r E_{\theta}, & -E_z\\
  E_r & 0 & r B_z & - B_{\theta}\\
  r E_{\theta} & -r B_z & 0 & r B_r\\
  E_z & B_{\theta} & -r B_r & 0
 \end{matrix}\right),
\quad \mbox{and} \quad
 G_{\mu\nu} = \left(
 \begin{matrix}
  0 & H_r & r H_{\theta}, & H_z\\
  -H_r & 0 & r D_z & -D_{\theta}\\
  -r H_{\theta} & -r D_z & 0 & r D_r\\
  -H_z & D_{\theta} & -r D_r & 0
 \end{matrix}\right).
\end{equation}
Matching the constitutive equation $\mathbf{G} = \boldsymbol{\chi}(\star\mathbf{F})$ for Minkowski space-time in cylindrical coordinates results in the identification
\small
\begin{equation} \label{Eq:CylindricalChi}
 \chi\indices{_{\gamma\delta}^{\sigma\rho}} = \frac12\left(
\begin{matrix}
 \left(\begin{smallmatrix}
 0 & 0 & 0 & 0\\ 0 & 0 & 0 & 0\\ 0 & 0 & 0 & 0\\ 0 & 0 & 0 & 0\\
 \end{smallmatrix} \right) &

* &

* &

* \\[15pt]

 \left(\begin{smallmatrix}
 0 & -\mu^{-1}_{rr} & -\tfrac{\mu^{-1}_{r\theta}}{r} & -\mu^{-1}_{rz}\\ 
 \mu^{-1}_{rr} & 0 & -\tfrac{\gamma_{1rz}}{r} & \gamma_{1r\theta}\\ 
 \tfrac{\mu^{-1}_{r\theta}}{r} & \tfrac{\gamma_{1rz}}{r} & 0 & -\tfrac{\gamma_{1rr}}{r}\\ 
 \mu^{-1}_{rz} & -\gamma_{1r\theta} & \tfrac{\gamma_{1rr}}{r} & 0\\
 \end{smallmatrix} \right) &

 \left(\begin{smallmatrix}
 0 & 0 & 0 & 0\\ 0 & 0 & 0 & 0\\ 0 & 0 & 0 & 0\\ 0 & 0 & 0 & 0\\
 \end{smallmatrix} \right) &

* &

* \\[20pt]

 \left(\begin{smallmatrix}
 0 & -r\mu^{-1}_{\theta r} & -\mu^{-1}_{\theta\theta} & -r\mu^{-1}_{\theta z}\\ 
 r\mu^{-1}_{\theta r} & 0 & -\gamma_{1\theta z} & r\gamma_{1\theta\theta}\\ 
 \mu^{-1}_{\theta\theta} & \gamma_{1\theta z} & 0 & -\gamma_{1\theta r}\\ 
 r\mu^{-1}_{\theta z} & -r\gamma_{1\theta\theta} & \gamma_{1\theta r} & 0\\
 \end{smallmatrix} \right) &

 \left(\begin{smallmatrix}
 0 & -r\gamma_{2zr} & -\gamma_{2z\theta} & -r\gamma_{2zz}\\
 r\gamma_{2zr} & 0 & -\epsilon_{zz} & r\epsilon_{z\theta}\\
 \gamma_{2z\theta} & \epsilon_{zz} & 0 & -\epsilon_{zr}\\
 r\gamma_{2zz} & -r\epsilon_{z\theta} & \epsilon_{zr} & 0\\
 \end{smallmatrix} \right) &

 \left(\begin{smallmatrix}
 0 & 0 & 0 & 0\\ 0 & 0 & 0 & 0\\ 0 & 0 & 0 & 0\\ 0 & 0 & 0 & 0\\
 \end{smallmatrix} \right) &

* \\[20pt]

 \left(\begin{smallmatrix}
 0 & -\mu^{-1}_{zr} & -\tfrac{\mu^{-1}_{z\theta}}{r} & -\mu^{-1}_{zz}\\ 
 \mu^{-1}_{zr} & 0 & -\tfrac{\gamma_{1zz}}{r} & \gamma_{1z\theta}\\ 
 \tfrac{\mu^{-1}_{z\theta}}{r} & \tfrac{\gamma_{1zz}}{r} & 0 & -\tfrac{\gamma_{1zr}}{r}\\ 
 \mu^{-1}_{zz} & -\gamma_{1z\theta} & \tfrac{\gamma_{1zr}}{r} & 0\\
 \end{smallmatrix} \right)  &

 \left(\begin{smallmatrix}
 0 & \gamma_{2\theta r} & \tfrac{\gamma_{2\theta\theta}}{r} & \gamma_{2\theta z}\\
 -\gamma_{2\theta r} & 0 & \tfrac{\epsilon_{\theta z}}{r} & -\epsilon_{\theta\theta}\\
 -\tfrac{\gamma_{2\theta\theta}}{r} & -\tfrac{\epsilon_{\theta z}}{r} & 0 & \tfrac{\epsilon_{\theta r}}{r}\\
 -\gamma_{2\theta z} & \epsilon_{\theta\theta} & -\tfrac{\epsilon_{\theta r}}{r} & 0\\
 \end{smallmatrix} \right)  &

 \left(\begin{smallmatrix}
 0 & -r\gamma_{2rr} & -\gamma_{2r\theta} & -r\gamma_{2rz}\\
 r\gamma_{2rr} & 0 & -\epsilon_{rz} & r\epsilon_{r\theta}\\
 \gamma_{2r\theta} & \epsilon_{rz} & 0 & -\epsilon_{rr}\\
 r\gamma_{2rz} & -r\epsilon_{r\theta} & \epsilon_{rr} & 0\\
 \end{smallmatrix} \right) &

 \left(\begin{smallmatrix}
 0 & 0 & 0 & 0\\ 0 & 0 & 0 & 0\\ 0 & 0 & 0 & 0\\ 0 & 0 & 0 & 0\\
 \end{smallmatrix} \right) \\
\end{matrix} \right).
\end{equation}
\normalsize
The form of the components in both of Eqs.\ (\ref{Eq:CylindricalF}) and (\ref{Eq:CylindricalChi}) could equally well be determined by transforming the Cartesian versions, Eqs.\ (\ref{Eq:FComponents}), (\ref{Eq:GComponents}), and (\ref{Eq:CartesianChi}), with the appropriate Cartesian-cylindrical transformation matrix.

In spherical coordinates $(t,r,\theta,\varphi)$, the Minkowski metric tensor has components $g_{\mu\nu} = diag(-1,1,r^2,r^2\sin^2\theta)$.  This time there are two dimensionless coordinates, so
\begin{equation}
 \mathbf{F} = E_r \mathrm{d}r\wedge\mathrm{d}t + r E_{\theta}\mathrm{d}\theta\wedge\mathrm{d}t +r \sin\theta E_{\varphi}\mathrm{d}\varphi\wedge\mathrm{d}t + r^2\sin\theta B_r\mathrm{d}\theta\wedge\mathrm{d}\varphi - r\sin\theta B_{\theta}\mathrm{d}r\wedge\mathrm{d}\varphi +  r B_{\varphi}\mathrm{d}r\wedge\mathrm{d}\theta
\end{equation}
(every $\mathrm{d}\theta$ gets a factor of $r$, every $\mathrm{d}\varphi$ gets a factor of $r\sin\theta$), which is written in matrix form as
\begin{equation} \label{Eq:SphericalF}
 F_{\mu\nu} = \left(
 \begin{matrix}
  0 & -E_r & -r E_{\theta}, & -r\sin\theta E_{\varphi}\\
  E_r & 0 & r B_{\varphi} & - r\sin\theta B_{\theta}\\
  r E_{\theta} & -r B_{\varphi} & 0 & r^2\sin\theta B_r\\
  r\sin\theta E_{\varphi} & r\sin\theta B_{\theta} & -r^2\sin\theta B_r & 0
 \end{matrix}\right).
\end{equation}
Letting $s=\sin\theta$, the same identification procedure as before leads to
\footnotesize
\begin{multline} \label{Eq:SphericalChi}
 \chi\indices{_{\gamma\delta}^{\sigma\rho}} \\ = \frac12\left(
\begin{matrix}
 \left(\begin{smallmatrix}
 0 & 0 & 0 & 0\\ 0 & 0 & 0 & 0\\ 0 & 0 & 0 & 0\\ 0 & 0 & 0 & 0
 \end{smallmatrix} \right) &
* &
* &
* \\[20pt]
 \left(\begin{smallmatrix}
 0 & -\mu^{-1}_{rr} & -\frac{\mu^{-1}_{r\theta}}{r} & -\frac{\mu^{-1}_{r\varphi}}{rs}\\
 \mu^{-1}_{rr} & 0 & \frac{-\gamma_{1r\varphi}}{r} & \frac{\gamma_{1r\theta}}{rs}\\ 
 \frac{\mu^{-1}_{r\theta}}{r} & \frac{\gamma_{1r\varphi}}{r} & 0 & \frac{-\gamma_{1rr}}{r^2s}\\
 \frac{\mu^{-1}_{r\varphi}}{rs} & \frac{-\gamma_{1r\theta}}{rs} & \frac{\gamma_{1rr}}{r^2s} & 0
 \end{smallmatrix} \right) &
 \left(\begin{smallmatrix}
 0 & 0 & 0 & 0\\0 & 0 & 0 & 0\\0 & 0 & 0 & 0\\0 & 0 & 0 & 0
 \end{smallmatrix} \right) &
*&
* \\[20pt]
 \left(\begin{smallmatrix}
 0 & -r \mu^{-1}_{\theta r} & -\mu^{-1}_{\theta\theta} & \frac{-\mu^{-1}_{\theta\varphi}}{s}\\ 
 r\mu^{-1}_{\theta r} & 0 & -\gamma_{1\theta\varphi}  & \frac{\gamma_{1\theta\theta}}{s}\\ 
 \mu^{-1}_{\theta\theta} & \gamma_{1\theta\varphi} & 0 & \frac{-\gamma_{1\theta r}}{rs}\\ 
 \frac{\mu^{-1}_{\theta\varphi}}{s} & \frac{-\gamma_{1\theta\theta}}{s} & \frac{\gamma_{1\theta r}}{rs} & 0
 \end{smallmatrix} \right) &
 \left(\begin{smallmatrix}
 0 & -r\gamma_{2\varphi r} & -\gamma_{2\varphi\theta} & \frac{-\gamma_{2\varphi\varphi}}{s}\\
 r\gamma_{2\varphi r} & 0 & -\epsilon_{\varphi\varphi} & \frac{\epsilon_{\varphi\theta}}{s}\\
 \gamma_{2\varphi\theta} & \epsilon_{\varphi\varphi} & 0 & \frac{-\epsilon_{\varphi}}{rs}\\
 \frac{\gamma_{2\varphi\varphi}}{s} & \frac{-\epsilon_{\varphi\theta}}{s} & \frac{\epsilon_{\varphi}}{rs} & 0
 \end{smallmatrix} \right) &
 \left(\begin{smallmatrix}
 0 & 0 & 0 & 0\\0 & 0 & 0 & 0\\0 & 0 & 0 & 0\\0 & 0 & 0 & 0
 \end{smallmatrix} \right) &
* \\[20pt]
 \left(\begin{smallmatrix}
 0 & -rs\mu^{-1}_{\varphi r} & -s\mu^{-1}_{\varphi\theta} & -\mu^{-1}_{\varphi\varphi}\\
 rs\mu^{-1}_{\varphi r} & 0 & -s\gamma_{1\varphi\varphi} & \gamma_{1\varphi\theta}\\
 s\mu^{-1}_{\varphi\theta} & s\gamma_{1\varphi\varphi} & 0 & \frac{-\gamma_{1\varphi r}}{r}\\
 \mu^{-1}_{\varphi\varphi} & -\gamma_{1\varphi\theta} & \frac{\gamma_{1\varphi r}}{r} & 0\\
 \end{smallmatrix} \right) &
 \left(\begin{smallmatrix}
 0 & rs\gamma_{2\theta r} & s\gamma_{2\theta\theta} & \gamma_{2\theta\varphi}\\
 -rs\gamma_{2\theta r} & 0 & s\epsilon_{\theta\varphi} & -\epsilon_{\theta\theta}\\
 -s\gamma_{2\theta\theta} & -s\epsilon_{\theta\varphi} & 0 & \frac{\epsilon_{\theta r}}{r}\\
 -\gamma_{2\theta\varphi} & \epsilon_{\theta\theta} & \frac{-\epsilon_{\theta r}}{r} & 0
 \end{smallmatrix} \right) & 
 \left(\begin{smallmatrix}
 0 & -r^2s\gamma_{2rr} & -rs\gamma_{2r\theta} & -r\gamma_{2r\varphi}\\
 r^2s\gamma_{2rr} & 0 & -rs\epsilon_{r\varphi} & r\epsilon_{r\theta}\\
 rs\gamma_{2r\theta} & rs\epsilon_{r\varphi} & 0 & -\epsilon_{rr}\\
 r\gamma_{2r\varphi} & -r\epsilon_{r\theta} & \epsilon_{rr} & 0
 \end{smallmatrix} \right) &
 \left(\begin{smallmatrix}
 0 & 0 & 0 & 0\\0 & 0 & 0 & 0\\0 & 0 & 0 & 0\\0 & 0 & 0 & 0
 \end{smallmatrix} \right)
\end{matrix} \right).
\end{multline}
\normalsize
Like the cylindrical result, the spherical result could be obtained equally well by transforming the Cartesian Eqs.\ (\ref{Eq:FComponents}), (\ref{Eq:GComponents}), and (\ref{Eq:CartesianChi}), with the appropriate Cartesian-spherical transformation matrix.
\section{Pullback and Pushforward} \label{Sec:Geometry}

Suppose two manifolds are related by a smooth map $\varphi:M\to N$.  Thus $\varphi$ maps the points of $M$ to an image in $N$, $p\in M \to \varphi(p)\in N$.  
Given a function $f$ on $N$, the \textit{pullback} of $f$ by $\varphi$, denoted $\varphi^*f$ is the composite function $f\circ \varphi$.  
So the pullback $\varphi^*$ takes a function $f$ on $N$ to a $\varphi$-related function $\varphi^*f$ on $M$.

The differential of $\varphi$ at a point $p\in M$ defines a map of tangent spaces $\mathrm{d}\varphi:T_pM\to T_{\varphi(p)}N$.  
Given a vector $V_p\in T_pM$, the \textit{pushforward} of $V_p$ assigns a $\varphi$-related vector $\mathrm{d}\varphi(V_p)\in T_{\varphi(p)}N$, such that
\begin{equation}
 \mathrm{d}\varphi(V)_{\varphi(p)}f = V_p(\varphi^*f)
\end{equation}
for some $f$ on $N$.  In particular, given coordinate charts $\{x^{\mu}\}$ on $M$ and $\{y^{\mu}\}$ on $N$ this is
\begin{alignat}{2}
 \left(\mathrm{d}\varphi(V_p)\right)^{\alpha}_{\varphi(p)}\frac{\partial}{\partial y^{\alpha}} f & = V^{\mu}_p\frac{\partial}{\partial x^{\mu}}(\varphi^*f) \\ &= V^{\mu}_p\frac{\partial}{\partial x^{\mu}}(f\circ\varphi) \\ &= V^{\mu}_p\left. \frac{\partial\varphi^{\alpha}}{\partial x^{\mu}}\right|_p\frac{\partial}{\partial y^{\alpha}}f.
\end{alignat}
Thus the pushforward of $V_p$ is
\begin{equation}
 \left(\mathrm{d}\varphi(V_p)\right)_{\varphi(p)}^{\alpha} = \Lambda\indices{^{\alpha}_{\mu}} V^{\mu}_p,
\end{equation}
where
\begin{equation}
 \Lambda\indices{^{\alpha}_{\mu}} = \left. \frac{\partial\varphi^{\alpha}}{\partial x^{\mu}}\right|_p
\end{equation}
is the Jacobian matrix of the map $\varphi$ evaluated at $p$.  
Note that the pushforward of a \textit{vector field} is not always defined.  
For example, if $\varphi$ is not injective then for two distinct points $p,q\in M$ such that $\varphi(p)=\varphi(q)\in N$, the vectors $V_p$ and $V_q$ would be pushed to the same point in $N$, but there is no reason why $\mathrm{d}\varphi(V_p)$ should equal $\mathrm{d}\varphi(V_q)$. 
However, if $\varphi$ is a diffeomorphism, then the pushforward of a vector field is well-defined.

Like a function (which is a 0-form), a 1-form on $N$ can be pulled back to a 1-form on $M$ such that
\begin{equation}
 \left(\varphi^*\omega\right)(V_p) = \omega\left(\mathrm{d}\varphi(V_p)\right).
\end{equation}
In particular, expressing this in coordinate charts as above, we find
\begin{equation}
 \left[\left.\left(\varphi^*\omega_{\varphi(p)}\right)\right|_p\right]_{\alpha} \mathrm{d}x^{\alpha}\left(V_p^{\beta}\frac{\partial}{\partial x^{\beta}}\right) = (\omega_{\varphi(p)})_{\alpha}\mathrm{d}y^{\alpha}\left(\left.\frac{\partial\varphi^{\beta}}{\partial x^{\mu}}\right|_pV_p^{\mu}\frac{\partial}{\partial y^{\beta}}\right).
\end{equation}
Which simplifies to
\begin{equation}
 \left[\left.\left(\varphi^*\omega_{\varphi(p)}\right)\right|_p\right]_{\alpha}V^{\alpha}= \left(\left.\frac{\partial\varphi^{\beta}}{\partial x^{\mu}}\right|_p \left(\omega_{\varphi(p)}\right)_{\alpha}\right) V^{\alpha},
\end{equation}
or
\begin{equation}
 (\varphi^*\omega)_{\alpha} = \omega_{\beta}\Lambda\indices{^{\beta}_{\alpha}},
\end{equation}
where $\Lambda\indices{^{\beta}_{\alpha}}$ is again the Jacobian matrix of $\varphi$ evaluated at $p$.


\end{document}